\documentclass{article}

\usepackage{arxiv}

\usepackage[utf8]{inputenc} 
\usepackage[T1]{fontenc}    
\usepackage{hyperref}       
\usepackage{url}            
\usepackage{booktabs}       
\usepackage{amsfonts}       
\usepackage{nicefrac}       
\usepackage{microtype}      
\usepackage{lipsum}
\usepackage{graphicx}
\graphicspath{ {./images/} }

\title{Integration of blockchain in smart systems: problems and opportunities for real-time sensor data storage}

\author{
 Naseem Alsadi \\
  Intelligent and Cognitive Engineering Laboratory\\
  McMaster University\\
  Hamilton, ON L8S 4L8 \\
  \texttt{alsadin@mcmaster.ca} \\
   \And
 Syed Zaidi \\
  Intelligent and Cognitive Engineering Laboratory\\
  McMaster University\\
  Hamilton, ON L8S 4L8 \\
  \And
 Mankaran Rooprai \\
  Intelligent and Cognitive Engineering Laboratory\\
  McMaster University\\
  Hamilton, ON L8S 4L8 \\
  \And
 Stephen A. Gadsden \\
  Intelligent and Cognitive Engineering Laboratory\\
  McMaster University\\
  Hamilton, ON L8S 4L8 \\
    \And
 John Yawney \\
  Adastra Corporation\\
  Thornhill, ON L3T 7M8\\
}

\begin{document}
\maketitle
\begin{abstract}
The internet of things (IoT) and other emerging ubiquitous technologies are supporting the rapid spread of smart systems, which has underlined the need for safe, open, and decentralized data storage solutions. With its inherent decentralization and immutability, blockchain offers itself as a potential solution for these requirements. However, the practicality of incorporating blockchain into real-time sensor data storage systems is a topic that demands in-depth examination. While blockchain promises unmatched data security and auditability, some intrinsic qualities, namely scalability restrictions, transactional delays, and escalating storage demands, impede its seamless deployment in high-frequency, voluminous data contexts typical of real-time sensors. This essay launches a methodical investigation into these difficulties, illuminating their underlying causes, potential effects, and potential countermeasures. 

\end{abstract}


\section{Introduction}
The advent of blockchain technology has paved the way for groundbreaking transformations across various sectors, heralding a new era of decentralization and security in digital transactions and data management. This innovative technology, best known for underpinning cryptocurrencies, extends its utility far beyond, promising to revolutionize not just the financial landscape but also the way we conceive and implement smart systems and the Internet of Things (IoT) [1 - 6]. By leveraging the intrinsic principles of decentralization, blockchain technology offers a robust framework for enhancing system efficiency, security, and transparency. However, the integration of blockchain within the domain of real-time sensor data storage presents a unique set of challenges that necessitate a nuanced understanding and innovative approaches. 

Real-time sensor networks, integral to IoT ecosystems, generate copious amounts of data pivotal for various applications ranging from environmental monitoring to smart cities. The seamless incorporation of blockchain technology into this realm promises enhanced security and integrity for sensor-derived data. Yet, the practicalities of implementing blockchain for real-time sensor storage unveil complexities related to data volume, speed of transactions, cost implications, data immutability, and storage scalability. These challenges underscore the tension between the theoretical ideals of blockchain technology and the pragmatic demands of high-frequency sensor environments. 

This paper delves into the multifaceted challenges that arise when deploying blockchain technology for real-time sensor data storage. It explores the intricacies of managing vast data volumes generated by sensors, the speed limitations imposed by consensus mechanisms in public blockchains, the economic considerations of frequent data transactions, the dilemma of data immutability versus the need for mutable storage, and the growing concerns around blockchain storage capacity. Addressing these challenges is crucial for harnessing the full potential of blockchain in enhancing the security and efficiency of IoT and smart systems. Consequently, the paper also proposes potential solutions and innovative approaches, including off-chain storage strategies, the utilization of sidechains, blockchain sharding, the adoption of private or consortium blockchains, and the development of more efficient consensus mechanisms. These solutions aim to reconcile the inherent properties of blockchain with the dynamic requirements of real-time sensor storage, paving the way for a more secure, efficient, and scalable integration of blockchain technology within IoT ecosystems.

\section{Challenges of Implementing Blockchain for Real-Time Sensor Storage}

The usage of blockchain as a solution to the issues faced by IoT systems is not without its own challenges. In this paper, we will go over a few of these hurdles in implementing blockchain for real-time sensor storage. Namely, these are the massive data volume, speed, cost implications, data immutability, and storage concerns of the technology.

\subsection{Data Volume}

High-frequency sensors, especially in clustered deployments, generate vast amounts of data. Storing this on a blockchain can strain the network, leading to scalability issues, especially in public blockchains. Because of blockchain’s key attribute of being immutable for security and verification purposes, the data generated by smart systems will only continuously increase the size of the blockchain. This can lead to data storage issues. The problem is compounded when we consider that smart systems employ the use of high-frequency sensors, further increasing the rate at which data is generated. As data comes in, it must be processed as a transaction and inputted into a block. Current popular public blockchains, such as Bitcoin and Ethereum, have relatively poor throughput with only 7 transactions per second (TPS) for Bitcoin and 20 for Ethereum [7]. High frequency sensors thus have the ability to potentially overwhelm public networks, resulting in delays for transaction processing and block confirmation. And as the amount of transactions grows, so does the need for storage. At this point, the blockchain trilemma is encountered where trade-offs are often made between security and decentralization for scalability [8]. IoT devices are meant to be cheap and light with small storage capacities. Storage requirements will eventually surpass the capabilities of IoT devices which, without cutting-edge solutions to this problem, might require the sacrifice of decentralization for a node or a group of nodes specially designed for increased storage capacity. Alternatively, the amount or quality of data could be reduced which would compromise security.

\subsection{Speed}

Real time sensor data storage requires rapid processing that is not often found in public blockchains. There are three main reasons why blockchains underperform traditional centralized databases when it comes to speed. The first is that the signature verification process is computationally complex. Couple this with the fact that it must be done for every transaction and it becomes easy to see why regular databases are often faster than blockchains since they only rely on the connection being secure to verify that a transaction is trustworthy [9]. When it comes to IoT devices, the low computational power of these devices further presents a challenge towards fast signature verification. Secondly, the consensus mechanism needed to reach trust across the network is also very resource intensive [10]. The Proof-of-Work (PoW) model used by Bitcoin, for instance, requires that each node find a hash value less than a specific number in order to add a block to the blockchain. However, this means that multiple nodes can find one such hash value at the same time, creating temporary forks where some nodes add on to the branch with a certain hash value and others to a different branch. Eventually the longest branch, or that with the most PoW, is appended to the blockchain and all other branches are thrown away [11]. These thrown away branches are examples of wasted time and computational resources. While such mechanisms may be secure, they require plenty of wasted time and high computational power, all slowing down the rate of transactions. This leads to the third point, redundancy. All transactions are processed independently by every node in the network all for the same end result [9]. Though this is necessary to ensure the trustworthiness of each block, it is still an example of extra computation required as opposed to databases that only process them once. The blockchain's requirement for continuous verification and trustworthiness checks introduces a level of computational overhead not typically present in databases.

\subsection{Cost Implications}

Given the storage requirements and advanced computation required for real-time sensor data storage in a blockchain, all within a compact consumer IoT device, it should be no surprise that implementing such requirements will require a hefty sum. If public blockchains are used to process transactions or use smart contracts, then fees will be incurred. These fees can vary based on the blockchain network and its congestion. Each sensor data entry requires a transaction, leading to a cumulative cost for storing data. Frequent transactions, especially in scenarios with a high sensor data volume, contribute to increased transaction costs. Blockchain's decentralized nature means that each node in the network must store a copy of the entire blockchain. Implementing mechanisms to optimize the handling of sensor data, such as aggregating or compressing data before storing it on the blockchain, can help reduce transaction volumes and associated costs. Smart contract logic can be designed to filter and store only essential data, mitigating the impact of high-frequency but less critical sensor readings. A study by Peker et al showed that the costs of using Ethereum for real-time sensor data storage was significantly reduced by implementing numerical data encoding on and off a smart contract [12]. Solving the slow computation issue within a compact IoT device follows the traditional hardware problem that the processor industry is constantly working towards - creating smaller processors while making them more powerful. This follows from Moore’s law where the primary issue is the number of transistors.

\subsection{Data Immutability}

Blockchain's primary strength lies in its ability to create a tamper-resistant and auditable history of transactions. Immutability is achieved by designating a consensus mechanism, like proof-of-work or proof-of-stake, to validate and agree on the addition of new blocks to the chain. Once a block is added, it becomes computationally infeasible to alter any previous blocks. Sensors generate a constant stream of data, and the need for quick, flexible responses to changing conditions may conflict with the immutability principle of blockchain. For instance, sensor data may need to be updated or corrected due to calibration errors, changing environmental conditions, or improved algorithms. Imposing immutability in such scenarios can be restrictive. In certain cases, regulations or privacy requirements may necessitate the removal of sensitive information from the system after a certain period. Immutability makes it challenging to comply with such data protection regulations as once data is recorded, it cannot be erased or anonymized easily. A recent article by Politou et al already outlines conflict between blockchain technologies and regulatory requirements due to the right to be forgotten [13]. The European Union’s recent General Data Protection Regulation would require that some data be removed in certain circumstances, which is in conflict with blockchain’s key attribute of being immutable. Redacting or updating data disrupts the consistency of data in an immutable blockchain, compromising the security guarantees it offers. This disruption can lead to unexpected vulnerabilities like double-spending attacks and forks in the blockchain network [14].

\subsection{Storage Concerns}

IoT devices and sensors are meant to be compact, limiting the amount of storage facilitating technology that can be implemented within. This means that to have a large storage capacity, the devices must carry the most expensive and data dense storage solutions. Storage concerns in the context of implementing blockchain for real-time sensor storage arise due to the inherent design principles of blockchain technology. While blockchain provides transparency, security, and immutability, its decentralized and append-only nature can pose challenges related to storage, particularly for real-time sensor data applications. For real-time sensor data applications, where data is generated continuously, the blockchain can expand rapidly, leading to significant storage requirements over time. In a decentralized blockchain network, each node is required to maintain a copy of the entire blockchain for validation and consensus purposes. As the blockchain grows, nodes need substantial storage capacity to store the complete transaction history. This can become impractical for nodes with limited storage resources. The ever-increasing size of the blockchain can then also result in scalability challenges, affecting the performance and efficiency of the network. Retrieving and validating data from a large blockchain may lead to delays, especially in real-time applications where quick access to data is crucial. Setting up a new node in the network will also take some time as the new node will have to download the previously existing blockchain ledger, which at the time of the node’s addition may have increased to insurmountable sizes [15]. 

\section{Potential Solution and Approaches}

\subsection{Off-chain Storage}

This research emphasizes the significance of off-chain storage in blockchain applications, particularly in the context of Internet of Things (IoT) data. The distinction between critical and non-critical data allows for the implementation of hybrid storage solutions, where critical, immutable records are stored on-chain, ensuring their security and permanence. On the other hand, bulk sensor data, which may be lots of data, is stored off-chain with references or cryptographic hashes stored on the blockchain [16]. 

The use of off-chain storage solutions such as BigChainDB and Hawk becomes crucial in scenarios where information needs verification through the blockchain without making the entire information available on-chain. The challenge lies in linking the physical storage location and the hash on the chain, and two methods, smart contracts and distributed hash tables (DHT), offer solutions [16].

Smart contracts, being programmable and unchangeable, are key for setting terms in agreements, like storage duration and costs. But there can be challenges, like relying on the storage provider's online status. To tackle this, Distributed Hash Tables (DHTs) provide an alternative by creating a decentralized network of storage with a central index on the chain, ensuring backup and fault tolerance [16]. In the realm of IoT, especially in the food supply chain (FSC) using a lot of IoT data for traceability [17], as blockchain nodes collect more data, off-chain storage becomes necessary. Here, the data's hash is stored on-chain, acting as a link to access off-chain data. Smart contracts and DHTs are suggested to link on-chain and physical storage, with a shared file system among nodes managing off-chain storage. This method not only deals with changing off-chain storage needs but also introduces the idea of data expiration for traceable goods [17].

The integration of a Distributed Hash Table (DHT) is discussed as a technology that supports fault-tolerant, decentralized storage based on key-value pairs. The scalability and flexibility of DHTs make them suitable for handling large volumes of IoT data in a distributed and decentralized manner. The mention of the Interplanetary File System (IPFS) as a DHT-based technology highlights its potential for secure and distributed data storage, emphasizing its relevance in applications that leverage both IPFS and blockchain [18].

The reference to off-chain data, using cryptographic hashes or references, enables the storage of large datasets off-chain while maintaining integrity. The off-chain storage of data is essential for accommodating large or dynamic datasets, and the smart contract acts as a bridge by linking on-chain execution with off-chain data. Additionally, considerations for access control and permissions in the smart contract add a layer of security and governance to the management of off-chain data [16].

\subsection{Side Chains}

Sidechains, as described in [19], are isolated secondary blockchains that serve as secondary blockchains connected to the main blockchain through a two-way peg. This mechanism enables bidirectional transfer of assets between the mainchain and the sidechain, addressing critical issues such as scalability, performance, privacy, and security in traditional blockchains. The challenges of implementing protocol changes in public blockchains due to their decentralized nature are alleviated by sidechains [19]. By allowing the offloading of transactions to sidechains, specific scenarios such as mobile crowdsensing and energy trading, as highlighted in [20], can benefit from more efficient processing.

Sidechains are connected to the mainchain in a parent-child relationship, allowing transactions to be executed in a private network, thus achieving off-chain scaling. Each sidechain can be tailored to specific scenarios, such as health care, environmental monitoring, and transportation management. The goal is to offload most mainchain traffic to attached sidechains, with efficient processing, low on-chain cost, and guaranteed security for cross-chain transactions [21].

In a building equipped with IoT devices configured to different sidechains, data can be efficiently shared and processed locally, resembling a Local Area Network [22]. The integration of SPV proof and smart contracts facilitates secure communication and synchronization across multiple floors, illustrating the potential of sidechains in meeting real-time data requirements in IoT scenarios [22]. To achieve this transfer, assets are moved by creating transactions on the first blockchain to lock assets, and then generating transactions on the second blockchain with cryptographic proofs of the correct lock, as explained in [23]. In the context of Bitcoin, sidechains transfer existing assets from the parent chain, preventing unauthorized coin creation, and relying on the parent chain to maintain asset security and scarcity. This process is crucial for ensuring the integrity of the entire blockchain network.

In the realm of IoT, where devices like electric vehicles engage in peer-to-peer energy trading [20], the secure and lightweight financial infrastructure provided by sidechains becomes crucial. Traditional centralized IoT infrastructures suffer from scalability issues and lack external audibility, while sidechains offer a decentralized alternative that ensures secure and auditable transactions. Additionally, merged mining, outlined in [23], enhances the capabilities of sidechains in the IoT space by enabling users to transfer funds securely between the main chain (e.g., Bitcoin) and sidechains. This transfer, coupled with waiting periods and security measures, ensures the integrity of the sidechain against potential reorganizations. In summary, sidechains offer a tailored solution to scalability issues in IoT applications, providing a decentralized and efficient framework for secure transactions and data sharing among interconnected devices.

\subsection{Sharding}

The existing throughput of 7 transactions per second (tps) in Bitcoin poses challenges for real-time applications, acting as a bottleneck for blockchain systems. Sharding addresses this by dividing the blockchain network into independent shards [24].

Each shard operates as a complete blockchain system, processing transactions or storing a subset of the network state. Multiple shards enable parallel transaction processing, significantly improving overall blockchain throughput. Cross-shard transactions introduce additional computing and communication overhead, impacting completion times [24].
Cross-shard transactions, involving more than one shard, introduce additional computing and communication overhead. The shard processing the cross-shard transaction must communicate with other shards involved, impacting completion times. Shards operate independently, each with their validators, consensus rules, and storage [25].
Sharding brings several benefits, as outlined in [25]:
\begin{itemize}
    \item Scalability: Increased transaction throughput.
    \item Improved Performance: Faster transaction confirmation times and reduced latency.
    \item Enhanced Privacy: Isolation of sensitive data within specific shards, improving privacy.
    \item Increased Network Decentralization: More participants can become validators.
    \item Flexibility: Enables customization and optimization of the blockchain network for specific use cases.
    \item Interoperability: Facilitates cross-shard communication and transactions, promoting interoperability between blockchain networks.
    \item Lower Storage Requirements: Reduces storage needs for each node, leading to lower storage costs.
    \item Accessibility: Makes blockchain networks more accessible to a broader range of users, including smaller devices or resource-constrained environments.
\end{itemize}
In the context of the Internet of Things (IoT) blockchain, sharding is particularly relevant due to the dynamic nature of IoT environments. Challenges arise from the unpredictable number and distribution of devices, requiring dynamic adjustment of sharding. The throughput of IoT sharding blockchain must be able to linearly scale to accommodate the increasing transaction data [26].

One way to tackle the scaling issue in IoT mentioned above, geographic location becomes a crucial characteristic for IoT blockchain nodes [26]. Sharding technologies designed for IoT blockchains consider the geographic location of nodes, aiming to group geographically adjacent devices into the same shard. This approach minimizes network overhead within the shard, improves consensus efficiency, and reduces cross-shard transactions.

All in all, sharding stands out as a pivotal solution to enhance blockchain scalability, offering benefits such as increased throughput, improved performance, enhanced privacy, and greater network decentralization. However, challenges persist, especially in dynamic environments like IoT, where careful consideration of factors like geographic location is essential to optimize the effectiveness of sharding.

\subsection{Private or Consortium Blockchains}

Private and consortium blockchains offer distinct advantages in terms of privacy, performance, and security compared to public blockchains. Privacy is a crucial aspect in blockchain networks, and public blockchains often face challenges in this regard due to the lack of robust privacy mechanisms. The traditional use of public keys as identifiers in public blockchains, like Bitcoin, provides only basic privacy and exposes users to potential privacy violations [1]. On the other hand, private blockchains restrict participation to a predefined or limited number of participants, eliminating the need for proof of work and mining. For instance, private and permissioned blockchains, such as IBM Hyperledger Fabric, ensure that only authorized individuals participate. This approach allows for better control over information sharing, reduces the likelihood of Byzantine behavior, and eliminates the need for expensive consensus protocols like Proof of Work (PoW) [27], [28].

Consortium blockchains combine features of both public and private blockchains [28]. They are governed by a group of organizations, providing advantages such as access control permissions, decentralized governance, low energy consumption, transaction confidentiality, high throughput, and enhanced security and scalability. The paper by Merlec et al. [29] further supports the advantages of consortium blockchains, highlighting their fault tolerance capability and protection against disturbances, even in the presence of malicious nodes.

Additionally, private blockchains developed by consortiums of organizations offer benefits such as controlled information sharing, better performance, and scalability. Consortium blockchains use a decentralized governance system for consensus, enabling efficient transactions with lower energy and computing resource usage [28].

One use case of consortium blockchains is in industrial IoT (IIoT), where they ensure data unforgeability. Consortium blockchains stop adversaries from pretending to be IIoT nodes and harming the network by using decentralized structures and digitally signed transactions. The encrypted raw data and the inability to forge audited and stored data add an extra layer of security to IIoT applications.

Moreover, consortium blockchains find practical applications in various industries, such as in the development of a unified and secure peer-to-peer energy trading system [30]. The decentralized and secure nature of consortium blockchains is leveraged to enable localized peer-to-peer electricity trading among plug-in hybrid electric vehicles, showcasing the adaptability of this technology in different industries.

In conclusion, private and consortium blockchains offer tailored solutions for specific use cases, providing enhanced privacy, performance, and security compared to public blockchains. These advantages make them suitable for applications ranging from secure energy trading systems to industrial IoT scenarios.

\subsection{Efficient Consensus Mechanisms}

Blockchain technology, initially introduced with Proof of Work (PoW) as its consensus mechanism, has evolved to address challenges related to transaction validation speed and energy consumption [31]. The foundational concept of blockchain, as outlined by Nakamoto in the Bitcoin whitepaper, involves a proof-of-work system. PoW relies on miners solving cryptographic puzzles to validate transactions and secure the network. However, PoW has faced criticism due to its time-consuming nature, resource-intensive mining processes, and the significant time required for transaction confirmations.

In addition to the drawbacks of PoW, a dominant 51\% control by any node in the network could potentially compromise the blockchain's integrity. To address these issues, there has been a shift towards PoS. PoS replaces miners with validators, offering advantages such as energy efficiency, enhanced security against 51

Building upon PoW and PoS, DPoS introduces a consensus algorithm based on voting elections. Elected representatives, similar to a democratic congress, participate in consensus and block generation. While DPoS enhances throughput and reduces latency, it faces challenges such as low enthusiasm among voting nodes and delayed handling of malicious nodes [33].

Consensus mechanisms have found applications beyond cryptocurrency. In the Internet of Things (IoT), researchers have explored blockchain-driven solutions with integrated consensus mechanisms to enhance traceability, decision support in supply chains, and effective data dissemination in industrial IoT environments [33 - 35].

Considering the resource constraints of IoT devices, PoW is deemed impractical due to its demand for substantial resources. In the Internet of Things (IoT), researchers have explored blockchain-driven solutions with integrated consensus mechanisms to enhance traceability, decision support in supply chains, and effective data dissemination in industrial IoT environments. Considering the resource constraints of IoT devices, PoW is deemed impractical due to its demand for substantial resources. This has prompted the exploration of alternative consensus mechanisms that are more suitable for IoT, addressing issues of applicability and security [36].

Efficient consensus mechanisms are critical for the evolution of blockchain technology. The transition from PoW to PoS, along with innovations like DPoS, not only improves the scalability and security of blockchain networks but also opens avenues for broader applications, particularly in the IoT space.

\newpage
\section{References}

[1]	N. Alsadi, S. A. Gadsden, and J. Yawney, “A cognitive dynamics framework for practical blockchain applications,” in Disruptive Technologies in Information Sciences VII, SPIE, Jun. 2023, pp. 25–34. doi: 10.1117/12.2664067.

[2]	N. Alsadi, A. Giuliano, S. A. Gadsden, and J. Yawney, “An adaptive approach to blockchain in smart system applications,” in Big Data V: Learning, Analytics, and Applications, SPIE, Jun. 2023, pp. 27–32. doi: 10.1117/12.2662231.

[3]	N. Alsadi et al., “An anomaly detecting blockchain strategy for secure IoT networks,” in Disruptive Technologies in Information Sciences VI, SPIE, May 2022, pp. 90–98. doi: 10.1117/12.2618301.

[4]	F. Chen, Z. Xiao, L. Cui, Q. Lin, J. Li, and S. Yu, “Blockchain for Internet of things applications: A review and open issues,” Journal of Network and Computer Applications, vol. 172, p. 102839, Dec. 2020, doi: 10.1016/j.jnca.2020.102839.

[5]	P. Cui, U. Guin, A. Skjellum, and D. Umphress, “Blockchain in IoT: Current Trends, Challenges, and Future Roadmap,” J Hardw Syst Secur, vol. 3, no. 4, pp. 338–364, Dec. 2019, doi: 10.1007/s41635-019-00079-5.

[6]	E. U. Haque, A. Shah, J. Iqbal, S. S. Ullah, R. Alroobaea, and S. Hussain, “A scalable blockchain based framework for efficient IoT data management using lightweight consensus,” Sci Rep, vol. 14, no. 1, p. 7841, Apr. 2024, doi: 10.1038/s41598-024-58578-7.

[7]	D. Khan, L. T. Jung, and M. A. Hashmani, “Systematic Literature Review of Challenges in Blockchain Scalability,” Applied Sciences, vol. 11, no. 20, Art. no. 20, Jan. 2021, doi: 10.3390/app11209372.

[8]	G. Del Monte, D. Pennino, and M. Pizzonia, “Scaling Blockchains Without Giving up Decentralization and Security.” arXiv, Jun. 04, 2020. doi: 10.48550/arXiv.2005.06665.

[9]	“Blockchain and Distributed Hash Table Technology in Decentralized Systems.” Accessed: Apr. 10, 2024. [Online]. Available: https://www.diva-portal.org/smash/record.jsf?pid=diva2\%3A1587002\&dswid=-8126

[10]	A. Odiljon and K. Gai, “Efficiency Issues and Solutions in Blockchain: A Survey,” in Smart Blockchain, M. Qiu, Ed., Cham: Springer International Publishing, 2019, pp. 76–86. doi: 10.1007/978-3-030-34083-4\_8.

[11]	D. Fullmer and A. S. Morse, “Analysis of Difficulty Control in Bitcoin and Proof-of-Work Blockchains,” in 2018 IEEE Conference on Decision and Control (CDC), Dec. 2018, pp. 5988–5992. doi: 10.1109/CDC.2018.8619082.

[12]	Y. Kurt Peker, X. Rodriguez, J. Ericsson, S. J. Lee, and A. J. Perez, “A Cost Analysis of Internet of Things Sensor Data Storage on Blockchain via Smart Contracts,” Electronics, vol. 9, no. 2, Art. no. 2, Feb. 2020, doi: 10.3390/electronics9020244.

[13]	E. Politou, F. Casino, E. Alepis, and C. Patsakis, “Blockchain Mutability: Challenges and Proposed Solutions,” IEEE Transactions on Emerging Topics in Computing, vol. 9, no. 4, pp. 1972–1986, Oct. 2021, doi: 10.1109/TETC.2019.2949510.

[14]	“Exploring the redaction mechanisms of mutable blockchains: A comprehensive survey - Zhang - 2021 - International Journal of Intelligent Systems - Wiley Online Library.” Accessed: Apr. 10, 2024. [Online]. Available: https://onlinelibrary.wiley.com/doi/10.1002/int.22502

[15]	F. Buccafurri, G. Lax, S. Nicolazzo, and A. Nocera, “Overcoming Limits of Blockchain for IoT Applications,” in Proceedings of the 12th International Conference on Availability, Reliability and Security, in ARES ’17. New York, NY, USA: Association for Computing Machinery, Aug. 2017, pp. 1–6. doi: 10.1145/3098954.3098983.

[16]	T. Hepp, M. Sharinghousen, P. Ehret, A. Schoenhals, and B. Gipp, “On-chain vs. off-chain storage for supply- and blockchain integration,” it - Information Technology, vol. 60, no. 5–6, pp. 283–291, Dec. 2018, doi: 10.1515/itit-2018-0019.

[17]	J. Rupasena, T. Rewa, K. T. Hemachandra, and M. Liyanage, “Scalable Storage Scheme for Blockchain-Enabled IoT Equipped Food Supply Chains,” in 2021 Joint European Conference on Networks and Communications \& 6G Summit (EuCNC/6G Summit), Jun. 2021, pp. 300–305. doi: 10.1109/EuCNC/6GSummit51104.2021.9482449.

[18]	M. Alizadeh, “Blockchain and Distributed Hash Table Technology in Decentralized Systems,” 2021, Accessed: Apr. 10, 2024. [Online]. Available: https://urn.kb.se/resolve?urn=urn:nbn:se:ltu:diva-86795

[19]	A. Singh, K. Click, R. M. Parizi, Q. Zhang, A. Dehghantanha, and K.-K. R. Choo, “Sidechain technologies in blockchain networks: An examination and state-of-the-art review,” Journal of Network and Computer Applications, vol. 149, p. 102471, Jan. 2020, doi: 10.1016/j.jnca.2019.102471.

[20]	C. Worley and A. Skjellum, “Blockchain Tradeoffs and Challenges for Current and Emerging Applications: Generalization, Fragmentation, Sidechains, and Scalability,” in 2018 IEEE International Conference on Internet of Things (iThings) and IEEE Green Computing and Communications (GreenCom) and IEEE Cyber, Physical and Social Computing (CPSCom) and IEEE Smart Data (SmartData), Halifax, NS, Canada: IEEE, Jul. 2018, pp. 1582–1587. doi: 10.1109/Cybermatics\_2018.2018.00265.

[21]	F. Gai, J. Niu, M. M. Jalalzai, S. A. Tabatabaee, and C. Feng, “A Secure Sidechain for Decentralized Trading in Internet of Things,” IEEE Internet of Things Journal, vol. 11, no. 3, pp. 4029–4046, Feb. 2024, doi: 10.1109/JIOT.2023.3300051.

[22]	C. E. Ngubo, P. McBurney, and M. Dohler, “Blockchain, IoT and Sidechains,” Jun. 2018. Accessed: Apr. 10, 2024. [Online]. Available: https://www.semanticscholar.org/paper/Blockchain\%2C-IoT-and-Sidechains-Ngubo-McBurney/b764a3276b93aeb3bbe07e3e9ca5730081285d6d

[23]	A. Back et al., “Enabling Blockchain Innovations with Pegged Sidechains”.

[24]	“Throughput-oriented associated transaction assignment in sharding blockchains for IoT social data storage - ScienceDirect.” Accessed: Apr. 10, 2024. [Online]. Available: https://www.sciencedirect.com/science/article/pii/S2352864822001171

[25]	“MQTT and blockchain sharding: An approach to user-controlled data access with improved security and efficiency - ScienceDirect.” Accessed: Apr. 10, 2024. [Online]. Available: https://www.sciencedirect.com/science/article/pii/S2096720923000337

[26]	C. Luo et al., “Fission: Autonomous, Scalable Sharding for IoT Blockchain,” in 2022 IEEE 46th Annual Computers, Software, and Applications Conference (COMPSAC), Jun. 2022, pp. 956–965. doi: 10.1109/COMPSAC54236.2022.00148.

[27]	A. M. Gergely and B. Crainicu, “RandAdminSuite: A New Privacy-Enhancing Solution for Private Blockchains,” Procedia Manufacturing, vol. 46, pp. 562–569, Jan. 2020, doi: 10.1016/j.promfg.2020.03.081.

[28]	C. Mohan, “State of Public and Private Blockchains: Myths and Reality,” in Proceedings of the 2019 International Conference on Management of Data, in SIGMOD ’19. New York, NY, USA: Association for Computing Machinery, Jun. 2019, pp. 404–411. doi: 10.1145/3299869.3314116.

[29]	M. M. Merlec, M. M. Islam, Y. K. Lee, and H. P. In, “A Consortium Blockchain-Based Secure and Trusted Electronic Portfolio Management Scheme,” Sensors, vol. 22, no. 3, Art. no. 3, Jan. 2022, doi: 10.3390/s22031271.

[30]	J. Kang, R. Yu, X. Huang, S. Maharjan, Y. Zhang, and E. Hossain, “Enabling Localized Peer-to-Peer Electricity Trading Among Plug-in Hybrid Electric Vehicles Using Consortium Blockchains,” IEEE Transactions on Industrial Informatics, vol. 13, no. 6, pp. 3154–3164, Dec. 2017, doi: 10.1109/TII.2017.2709784.

[31]	S. Nakamoto, “Bitcoin: A Peer-to-Peer Electronic Cash System”.

[32]	B. Sriman,  s Kumar, and S. Prabakaran, “Blockchain Technology: Consensus Protocol Proof of Work and Proof of Stake,” 2020, pp. 395–406. doi: 10.1007/978-981-15-5566-4\_34.

[33]	Q. Hu, B. Yan, Y. Han, and J. Yu, “An Improved Delegated Proof of Stake Consensus Algorithm,” Procedia Computer Science, vol. 187, pp. 341–346, Jan. 2021, doi: 10.1016/j.procs.2021.04.109.

[34]	Y. P. Tsang, K. Choy, C.-H. Wu, G. T. s Ho, and H. Lam, “Blockchain-Driven IoT for Food Traceability With an Integrated Consensus Mechanism,” IEEE Access, vol. 7, pp. 129000–129017, Sep. 2019, doi: 10.1109/ACCESS.2019.2940227.

[35]	R. Huang, X. Yang, and P. Ajay, “Consensus mechanism for software-defined blockchain in internet of things,” Internet of Things and Cyber-Physical Systems, vol. 3, pp. 52–60, Jan. 2023, doi: 10.1016/j.iotcps.2022.12.004.

[36]	L. Qi, J. Tian, M. Chai, and H. Cai, “LightPoW: A trust based time-constrained PoW for blockchain in internet of things,” Computer Networks, vol. 220, p. 109480, Jan. 2023, doi: 10.1016/j.comnet.2022.109480.

\end{document}